\documentclass[%
 reprint,
 superscriptaddress,
 amsmath,amssymb,
 aps,
]{revtex4-1}

\usepackage[utf8]{inputenc}
\usepackage{graphicx}
\usepackage{dcolumn}
\usepackage{bm}
\usepackage{hyperref}
\usepackage{nicefrac}
\usepackage{bbm}
\usepackage{todonotes}

\newcommand{\ba}{\mathbf{a}}
\newcommand{\bD}{\mathbf{D}}
\newcommand{\bU}{\mathbf{U}}
\newcommand{\bb}{\mathbf{b}}
\newcommand{\ha}{\mathbf{a}}

\newcommand{\hc}{\mathbf{c}}
\newcommand{\hn}{\mathbf{N}}
\newcommand{\hph}{\bm{\varphi}}
\newcommand{\hth}{\bm{\theta}}
\newcommand{\hH}{\mathbf{H}}

\newcommand{\red}[1]{\textcolor{black}{#1}}
\newcommand{\bn}{\mathbf{N}}
\newcommand{\bph}{{\bm \varphi}}
\newcommand{\hHs}{\mathbf{H}_{\text{shunt}}}

\newcommand{\tha}{\widetilde{\mathbf{a}}}
\newcommand{\thb}{\widetilde{\mathbf{b}}}
\newcommand{\thn}{\widetilde{\mathbf{N}}}

\newcommand{\thth}{\widetilde{\bm{\theta}}}
\newcommand{\thH}{\widetilde{\mathbf{H}}}
\newcommand{\thHs}{\widetilde{\mathbf{H}}_{\text{shunt}}}

\newcommand{\ep}{A_p}
\renewcommand{\wp}{\omega_p}
\newcommand{\nbar}{\bar{n}}

\newcommand{\bra}[1]{\left\langle #1 \right|}
\newcommand{\ket}[1]{\left| #1 \right\rangle}

\newcommand{\braket}[2]{\left\langle #1 \right|\left. #2 \right\rangle}
\begin{document}


\title{\red{Structural instability of driven Josephson circuits prevented by an inductive shunt}}

\author{Lucas Verney}
\affiliation{QUANTIC team, INRIA de Paris, 2 Rue Simone Iff, 75012 Paris, France}
\affiliation{Laboratoire Pierre Aigrain, Ecole Normale Supérieure, PSL Research University, CNRS, Université Pierre et Marie Curie, Sorbonne Universités, Université Paris Diderot, Sorbonne Paris-Cité, 24 rue Lhomond, 75231 Paris Cedex 05, France}
\author{Raphaël Lescanne}
\affiliation{Laboratoire Pierre Aigrain, Ecole Normale Supérieure, PSL Research University, CNRS, Université Pierre et Marie Curie, Sorbonne Universités, Université Paris Diderot, Sorbonne Paris-Cité, 24 rue Lhomond, 75231 Paris Cedex 05, France}
\affiliation{QUANTIC team, INRIA de Paris, 2 Rue Simone Iff, 75012 Paris, France}
\author{Michel H. Devoret}
\affiliation{Department of Applied Physics, 15 Prospect St, Yale University, New Haven, CT 06511, USA}
\author{Zaki Leghtas}
\affiliation{Centre Automatique et Syst\`emes, Mines-ParisTech, PSL Research University, 60, bd Saint-Michel, 75006 Paris, France}
\affiliation{Laboratoire Pierre Aigrain, Ecole Normale Supérieure, PSL Research University, CNRS, Université Pierre et Marie Curie, Sorbonne Universités, Université Paris Diderot, Sorbonne Paris-Cité, 24 rue Lhomond, 75231 Paris Cedex 05, France}
\affiliation{QUANTIC team, INRIA de Paris, 2 Rue Simone Iff, 75012 Paris, France}
\author{Mazyar Mirrahimi}
\affiliation{QUANTIC team, INRIA de Paris, 2 Rue Simone Iff, 75012 Paris, France}
\affiliation{Yale Quantum Institute, Yale University, New Haven, CT 06520, USA}

\date{\today}

\begin{abstract}

\red{Superconducting circuits are a versatile platform to implement a multitude of Hamiltonians which  perform quantum computation, simulation and sensing tasks. A key ingredient for realizing a desired Hamiltonian is the irradiation of the circuit by a strong drive. These strong drives provide an in-situ control of couplings, which cannot be obtained by near-equilibrium Hamiltonians. However, as shown in this paper, out-of-equilibrium systems are easily plagued by complex dynamics leading to instabilities. Predicting and preventing these instabilities is crucial, both from a fundamental and application perspective.}
 \red{We propose an inductively shunted transmon as the elementary circuit optimized for strong parametric drives. Developing a novel numerical  approach that avoids the built-in limitations of perturbative analysis, we demonstrate that adding the inductive shunt significantly extends  the range of pump powers over which the circuit behaves in a stable manner. } 

\end{abstract}

\pacs{Valid PACS appear here} 
\maketitle

\section{Introduction}\label{sec:intro}
Josephson junctions are ideal non-dissipative elements that  realize nonlinear Hamiltonians for superconducting quantum circuits. {Compared to nonlinear crystals in the optical regime, Josephson circuits have a much larger ratio between multi-wave mixing and decoherence rates ~\cite{Wallraff2004,Schuster2007,Kirchmair2013}}. By applying off-resonant drives (pumps) verifying  frequency matching conditions, one can engineer various  Hamiltonians that are not obtainable statically. {This so-called parametric method  has been used, for instance, to achieve frequency conversion~\cite{Abdo2013}, quantum-limited amplification~\cite{CastellanosBeltran2008}, two-mode squeezing~\cite{Flurin2012}, transverse readout of a qubit~\cite{Vool2016}, and multi-photon exchanges between two modes~\cite{Leghtas2015}.  } \red{In all these applications, the rates of the engineered parametric couplings scale with the pump power. However, as observed in~\cite{Leghtas2015,Gao2018,Lescanne2018}, this scaling can be strongly limited by effects such as the induced deterioration of the coherence properties. }


\red{In this paper, we explain these limitations by analyzing the structural stability of the underlying dynamical system. We call a dynamical system structurally stable if small modifications of the parameters, such as the strength of the pumping drives, lead to small changes in its qualitative behavior, such as the asymptotic steady states of the driven-dissipative system. We show that the  ubiquitous system consisting of a transmon~\cite{Koch2007,Paik2011} coupled to a  cavity mode displays strong instabilities in this sense.} We predict that above a critical pump power the transmon state  escapes the Josephson potential confinement and {is sent to free-particle-like states. The circuit behaves then as if we had removed the junction, and this explains the jump of the cavity frequency towards its  bare (undressed) value, a phenomenon observed and used in the past for single shot qubit readout~\cite{Reed2010}. } \red{Next, to prevent the instability caused by this  escape from the {confining} potential, we propose to shunt} the transmon with an inductance smaller than the kinetic inductance of the junction. We show that, as a result of the additional harmonic confinement, this system behaves in a stable manner over a wide range of pump strengths. 

{\red{Non-perturbative numerical simulations} of these strongly driven nonlinear systems is particularly {challenging}. It requires simulating a master equation over a Hilbert space of large dimension, and with time-scales separated by many orders of magnitude~\cite{Faou2012}. \red{Here, we treat the dimension problem by performing transformations that displace correctly the high excitation manifold into a tractable one} (see Appendix~\ref{sec:models}). Also, usually, to simplify the dynamics, one starts by removing the fast time scales through rotating-wave approximations. However,  reliable simulations in the presence of strong drives require taking into account the counter-rotating terms in the Hamiltonian, whose importance have  been previously noticed by~\cite{Sank2016,Pietikainen-17}. }  Here, {we  avoid  time-averaging  the driven Hamiltonian, by using  the Floquet-Markov theory~\cite{Grifoni1998} to characterize the asymptotic behavior of the system.} {The periodically driven Hamiltonian of a general circuit, subject to  a single pump at frequency $\omega_p$ (Fig.~\ref{fig:main}(a) and (d)),  can be expressed in the Floquet states basis.} These Floquet states {$\{\ket{\Psi_\alpha(t)}\}_{\alpha}$}, corresponding to $2\pi/\omega_p$-periodic orbits of the system, are the eigenstates of the time-dependent Hamiltonian associated to eigenvalues {$\{\epsilon_\alpha\}_{\alpha}$} that are called the Floquet quasi-energies.  For any Markovian bath, and in the weak coupling limit, one achieves an effective Floquet master equation for the evolution of the open quantum system. In the absence of resonances~{\cite[Section 9.3]{Grifoni1998}}, this  Floquet master equation {admits, as the steady state,} a limit cycle of period $2\pi/\wp$ given by {$\rho_{ss}(t)=\sum_\alpha p_\alpha \ket{\Psi_\alpha(t)}\bra{\Psi_\alpha(t)}$}, a statistical mixture of Floquet states. The populations of these states are calculated through an extension of the Fermi golden rule to time-periodic systems~\cite{Grifoni1998} (also see Appendix~\ref{sec:floquet}).

\section{Driven transmon and structural instability}\label{sec:unshunted}
\begin{figure*}
    \includegraphics[width=\textwidth]{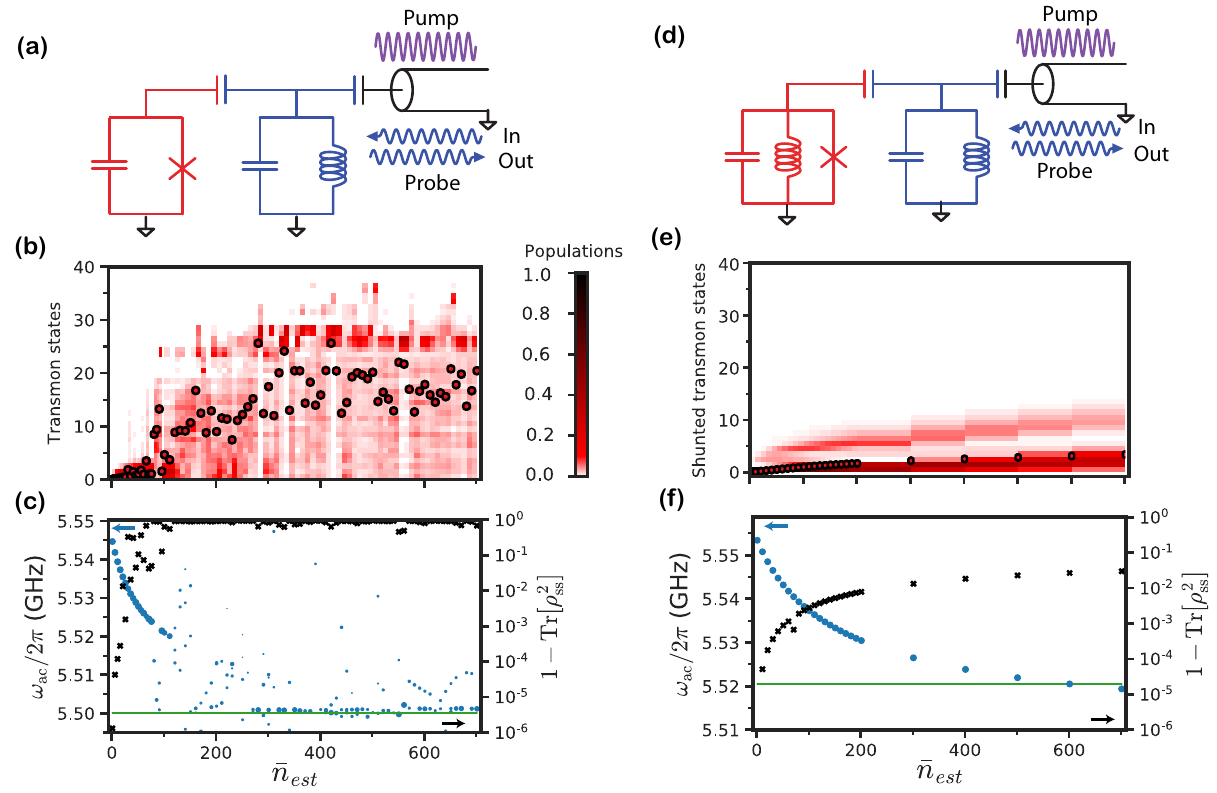}
    \caption{Floquet-Markov simulations (asymptotic regime) of the un-shunted and shunted transmon. (\textbf{a} and \textbf{d}) Circuits of a regular transmon and  an inductively shunted transmon, coupled to a harmonic oscillator{, and capacitively coupled to a transmission line.} A (strong) off-resonant microwave drive at frequency $\omega_p$, called a pump, is sent to the system through the transmission line.   (\textbf{b} and \textbf{e}){Populations of the transmon eigenstates $\ket{\eta_k}$, and shunted transmon states  $\ket{\nu_k}$, in the steady state $\rho_{ss}(0)$ of~\eqref{eq:transmon_int}-\eqref{eq:bath1} and of~\eqref{eq:shunted_displaced}-\eqref{eq:bath1}, as a function of pump power. In the un-shunted case, the parameters are taken to be  ${E_C/h = 150}$ MHz, ${E_J/h = 20}$ GHz, ${g/2\pi = 140}$ MHz, ${\omega_a/2\pi = 5.5 \text{GHz}}$ and ${\omega_p/2\pi = 6}$ GHz. For the shunted transmon, we use the same parameters except for ${E_J/h = 6}$ GHz and ${E_L/h = 14}$ GHz (leading to the same bare transmon frequency). Here, $\bar n_{\text{est}}=|\ep|^2/4{\left|\wp - \omega_a\right|^2}$ is an estimation of the circulating photon number $\nbar$, where we have used the bare oscillator frequency instead of the dressed one.}  {Red dots} indicate the average number of excitations in the transmon mode.  (\textbf{c} and \textbf{f})  Blue dots (left axis) correspond to the AC Stark shifted frequencies of the oscillator as a function of the pump power.  The areas of the points are proportional to the associated transition probabilities (see Appendix~\ref{sec:floquet}). Green horizontal line corresponds to the oscillator bare frequency $\omega_a$ in the un-shunted case, and to the renormalized frequency $\widetilde{\omega}_a$ in the shunted one. In the first case, at many pump powers, we observe multiple resonance frequencies corresponding to different transitions from the limit cycle to Floquet states. In contrast, in the second one, the frequency is unique and well-defined for all pump frequencies. This is also reflected by the impurity of the steady state (black crosses, right axis). While in the first case, the steady state is very mixed even for small pump strengths, in the second one, the impurity remains smaller than $3\%$.}
    \label{fig:main}
\end{figure*}

We start by considering   a  transmon  coupled to a harmonic oscillator (referred in the following simply as ``oscillator''). The Hamiltonian of this circuit (shown in Fig.~\ref{fig:main}(a)) is given by
\begin{equation}\label{eq:transmon_initial}\begin{split}
    \hH\left(t\right) = &\, \hbar \omega_a\, \ha^{\dag} \ha + 4 E_C\, \hn^2 %
        - E_J \cos\left(\bm{\hth}\right) \\
    & +i\hbar g\,\hn  \left(\ha^\dag - \ha\right) + %
        i\hbar \mathcal{A}_p(t)\left(\ha^\dag - \ha\right).
\end{split}\end{equation}

Here, $\hn$ and $\cos\left(\bm{\hth}\right)$ are the transmon mode operators corresponding to the number of Cooper pairs and their transfer across the junction, while $\ha$ and $\ha^{\dag}$ are photon annihilation and creation operators of the oscillator. We note that here the phase $\hth$ takes its values in the interval $[0,2\pi]$ and only periodic operators such as ${\cos(\bm{\hth})=(\sum_N \ket{N}\bra{N+1}+\text{h.c.})/2}$ are well-defined (here $\ket{N}$ are the charge states)~\cite{DevoretHouches}.  Furthermore, $E_C$ is the charging energy, $E_J$ is the Josephson {coupling} energy, $\omega_a$ is the bare frequency of the oscillator in absence of coupling to the transmon, $g$ is the coupling rate between the two modes. The pump is described by ${\mathcal{A}_p(t) = \ep\cos\left(\wp t\right)}$ with an amplitude $\ep$ and a frequency $\wp$ far detuned from the resonance frequencies of the system. Throughout this paper, we will consider as the basis, the tensor products of the oscillator Fock states $\left\{\ket{n}\right\}_{n=0}^\infty$ and the transmon states $\left\{\ket{\eta_k}\right\}_{k=0}^\infty$ (eigenstates of the transmon Hamiltonian $4 E_C \hn^2 - E_J \cos\left(\bm{\hth}\right)$). We model the dissipation as a capacitive coupling of the oscillator to a transmission line~\cite{GardinerZollerQuantumNoise}
\begin{equation}\label{eq:bath1}
    \hH_{SB}=\sum_{k}\hbar\omega_k\hc^\dag[\omega_k]\hc[\omega_k]-\hbar\Omega[\omega_k](\ha^\dag-\ha)(\hc^\dag[\omega_k]-\hc[\omega_k]).
\end{equation}
{Here the modes $\hc[\omega_k]$ are the bath modes and $\Omega[\omega_k]$ represents their coupling strengths to  the mode $\ha$. }

 We investigate the dynamics of this system for large pump amplitudes where the circulating photon number, given by  ${\nbar = {\left|\ep\right|^2}/{4 \left|\Delta_p\right|^2}}$ (with $\Delta_p$ the detuning between the pump frequency and the dressed oscillator frequency), can reach a few thousands. In order to {reduce the required truncation of the Hilbert space}, we consider a change of variables which takes into account such a coherent displacement of the oscillator. As shown in Appendix~\ref{sec:models}, the new Hamiltonian is given by
\begin{equation}\label{eq:transmon_int}\begin{split}
    \thH\left(t\right) &= \,\hbar \omega_a\, \tha^{\dag} \tha + 4 E_C\, \thn^2 \\ %
     &   - E_J \cos\left(\bm{\thth} + \xi\sin(\wp t) \right)
     +i\hbar g\,\thn \left(\tha^\dag - \tha\right),
\end{split}\end{equation}
where $\xi = 2g \omega_a \ep / \left[\wp \left( \omega_a^2-\wp^2\right)\right]$.

We have performed  Floquet-Markov-type simulations~\footnote{Simulations were run on a workstation with Python 3.5.2 and a modified version of QuTiP 4.2.0~\cite{qutip}.} assuming a white noise spectrum for the bath. In Fig.~\ref{fig:main}(b), we plot the populations of the transmon eigenstates $\left\{\ket{\eta_k}\right\}_{k=0}^{\infty}$ in ${\rho_{ss}(0)}$ as a function of pump power. {We do not plot the populations in the mode $\tha$, as the displaced oscillator remains close to its ground state. This confirms that the actual state is well-approximated by a coherent state as calculated in Appendix~\ref{sec:models}.} The dynamics of the displaced transmon mode exhibits two regimes. For ${\nbar\lesssim 100}$, the state remains pure  {(impurity given by the black crosses, right axis, in Fig.~\ref{fig:main}(c))} close to the ground state, except for a few pump power values. For ${\nbar \gtrsim 100}$ it rapidly turns into a mixed state of high number of excitations, above the cosine confinement. Indeed, the number of confined states is roughly given by the ratio between the depth of the cosine potential ($2E_J$) and the level spacings ($\approx \sqrt{8E_JE_C}$)~\cite{Koch2007}. With the parameters used in Fig.~\ref{fig:main}, we obtain about 8 confined levels.

Inspired by the experiments on the AC Stark shift~\cite{Schuster2005,Ong2011}, we simulated an excitation spectroscopy of such a driven system near the oscillator bare frequency. Each Floquet state {$\ket{\Psi_\alpha(t)}$, with a non-zero population in the  steady state $\rho_{ss}(t)$, can be excited to other Floquet states $\ket{\Psi_\beta(t)}$ by a weak probe drive at the frequency given by the difference of their quasi-energies $(\epsilon_\beta-\epsilon_\alpha)/\hbar$~\cite{Silveri2013,Pietikainen-17,Pietikainen-18} (see also Appendix~\ref{sec:floquet})}. In Fig.~\ref{fig:main}(c), we plot all these resonance frequencies as a function of the pump power. For each pump power, we may observe a few resonance frequencies corresponding to various transitions and various Floquet states populated in the limit cycle. For weak drives ${\nbar\ll 100}$, we observe a linear behavior in agreement with the usual AC Stark shift experiments~\cite{Schuster2005,Ong2011} and the associated theoretical work~\cite{Gambetta2006}. The behavior remains rather smooth up to ${\nbar\approx 100}$ with a slight curvature representing the effect of higher order nonlinearities~\cite{Vlastakis2013}. For ${\nbar\gtrsim 300}$ the dominant resonance frequency shifts near oscillator bare frequency. This can be physically understood by the fact that high-energy transmon states (energy above $2 E_J$) are not affected by the cosine potential and therefore are well approximated by charge states. When reaching these levels (Fig.~\ref{fig:main}(b)), the transmon mode acts as a free particle (similar to the ionization of an atom), whose dynamics follows that of the oscillator. The oscillator does no longer inherit a non-linearity from the transmon mode as evidenced by the jump of its resonance frequency towards the bare frequency $\omega_a$. These two regimes slightly overlap in the middle region (${100\lesssim \nbar\lesssim 300}$) which presents many transition frequencies.

Previously, such a jump in the resonance frequency has been observed in a setup with a single strong probe drive, and used to perform single shot measurements of the transmon qubit~\cite{Reed2010}. Various theoretical work have investigated this phenomenon assuming two-level~\cite{Bishop2010}, multi-level~\cite{Boissonneault2010,Mavrogordatos2017,Pietikainen-17}, and Duffing approximations~\cite{Elliott2016} of the transmon mode. In contrast to these approaches, the above numerical simulations of the full model \eqref{eq:transmon_int}-\eqref{eq:bath1}, and the experimental observations of~\cite{Lescanne2018}, illustrate that such a jump in the resonance frequency coincides with the excitation of the transmon mode to high energy levels well beyond the confinement potential.

\section{Inductively shunted transmon: a solution to instability}\label{sec:shunted}
The above analysis illustrates that in the parametric construction of a nonlinear Hamiltonian (such as the two-photon exchange between two modes), we are strongly limited in the  span of the pump strength. Above a critical threshold, the ionized transmon no longer induces any nonlinearity on the oscillator. Such a limitation has been observed through the heating of the transmon mode in~\cite{Gao2018}. Further confinement of the nonlinear mode should provide a larger span of exploitable pump strength. We propose here to shunt the transmon circuit with an inductance providing a harmonic confinement of the phase across the junction~\cite{Koch2009,Braumuller-16,Richer-17}.
The Hamiltonian of such a circuit (shown in Fig.~\ref{fig:main}(d)) is given by
{
\begin{equation}\label{eq:shunted_initial}\begin{split}
 &   \hHs\left(t\right) = \, \hbar \omega_a \, \ha^{\dag} \ha + 4 E_C\, \hn^2 %
        + \frac{E_L}{2}\, \hph^2 \\
    & - E_J \cos\left(\hph\right) +i\hbar g\,\hn \left(\ha^{\dag} - \ha\right)  %
        +i\hbar \mathcal{A}_p(t) \left(\ha^{\dag} - \ha\right).
\end{split}\end{equation}}
where $E_L$ is the shunt inductance energy and $\hph$ represents the \red{dimensionless} flux operator across the junction~\cite{DevoretHouches}. Previously, the inductively shunted Josephson junctions have been considered as superconducting qubit designs~\cite{Mooij1999,Manucharyan2009}. Here, we consider parameters comparable to a flux qubit ${E_C\ll E_J\lesssim E_L}$. However, rather than the coherence properties of this circuit, we are interested in its behavior as a nonlinear device in the strong pumping regime. While at large numbers of excitations, the harmonic potential $ \frac{E_L}{2}\, \hph^2$  dominates the nonlinear part $E_J \cos\left(\hph\right)$, the passage to the linear regime should be smoother than with the transmon. We therefore expect to be able to explore the nonlinearity up to a higher number of excitations.

Similarly to the un-shunted case, after a unitary transformation provided in Appendix~\ref{sec:models}, the Hamiltonian of the inductively shunted transmon becomes
\begin{multline}\label{eq:shunted_displaced}
    \thHs(t) =  \hbar \widetilde{\omega}_a \, \tha^{\dag}\tha + %
	    \hbar \widetilde{\omega}_b \, \thb^{\dag}\thb  \\
     - E_J \cos \left[\varphi_a^0\left(\tha + \tha^{\dag}\right) + \varphi_b^0 \left(\thb + \thb^{\dag}\right) +%
        \xi \sin\left(\omega_p t\right)\right]
\end{multline}
where $\widetilde{\omega}_a$ and $\widetilde{\omega}_b$ are renormalized frequencies, $\varphi_a^0$ and $\varphi_b^0$ are zero-point fluctuations of the two modes as seen by the Josephson junction and $\xi$ is a renormalized pump amplitude. Here the mode $\tha$ is closer to the initial oscillator mode $\ha$ and  the mode $\thb$ is closer to the  junction mode ($\varphi_a^0\ll \varphi_b^0$). In contrast to the un-shunted case, this change of variables ensures that both modes remain close to their ground state.  This is a direct consequence of the harmonic confinement and will be confirmed through numerical simulations.

We use again the Floquet-Markov framework to carry out the numerical simulations of the driven dissipative system~\eqref{eq:shunted_displaced} and \eqref{eq:bath1}. {While the calculations are done in the basis of the Fock states of the two modes $\tha$ and $\thb$, we plot the results in the shunted transmon basis $\{\ket{\nu_k}\}_{k=0}^\infty$ (eigenstates of the Hamiltonian $4E_C \hn^2+E_L\hph^2/2-E_J\cos(\hph)$)}. In Fig.~\ref{fig:main}(e), we plot the populations of the states $\ket{\nu_k}$ in the steady state together with its average number of excitations (red {dots}). We have not plotted the populations in the mode $\tha$, as it remains very close to its ground state. {We observe that the state $\rho_{ss}(0)$ follows a very smooth behavior and as shown in  Fig.~\ref{fig:main}(f), the impurity of $\rho_{ss}$ (black crosses, right axis) remains close to zero. As shown in Appendix~\ref{sec:floquet}, in the frame corresponding to $\tha$ and $\thb$, this steady state remains very close to the ground state for all values of the pump power.} Finally, Fig.~\ref{fig:main}(f) also illustrates the AC Stark shifted frequency of the resonator mode which is now well-defined for all values of the pump power.  {The simulation parameters are chosen such that the bare frequencies, impedances and coupling of the harmonic oscillator and the transmon mode coincide with those of the un-shunted case.  The important change concerns the dilution of the nonlinearity by the addition of the harmonic shunt with an energy $E_L$, about a factor of 2 larger than $E_J$ (see Appendix~\ref{sec:floquet} for simulations with other parameters and comments on the choice of factor 2).}

\begin{figure}[h!]
    \includegraphics[width=.95\columnwidth]{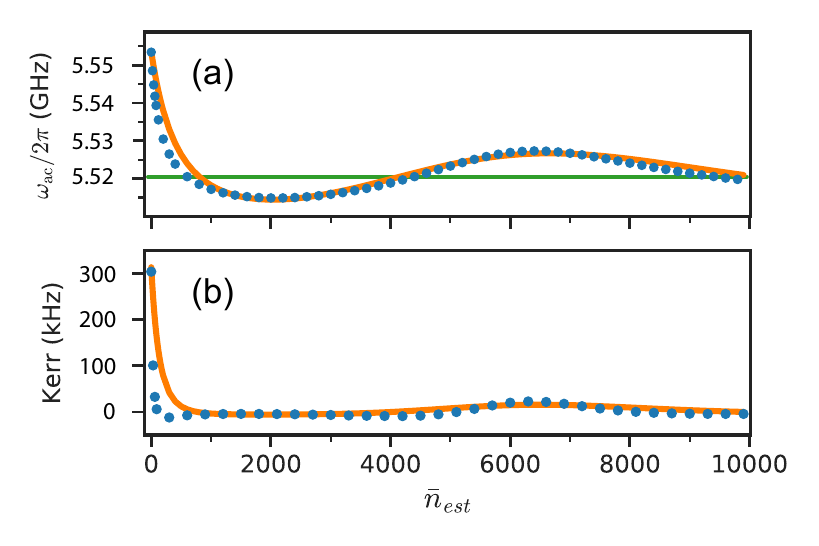}
    \caption{(a) Floquet-Markov simulations  of the AC Stark shifted frequencies, for very large cavity photon numbers in the shunted case. {The superimposed orange curve corresponds to the resonance frequency from the time-averaged model. Indeed, it is calculated as the difference between the two eigen-energies of the Hamiltonian~\eqref{eq:av} associated to the  dressed $\tha$-mode. The green  line corresponds to the oscillator renormalized frequency $\widetilde{\omega}_a$. (b) Strength of the induced Kerr of the most linear mode $\tha$, defined as the difference of the transition frequencies for the first and the second excitations. The orange curve corresponds to the expected induced Kerr strength from the time-averaged model~\eqref{eq:av}. }}
    \label{fig:extended}
\end{figure}

{As a result of this smooth behavior, we can extend the study to much higher pump powers. As illustrated in Fig.~\ref{fig:extended}(a), the AC Stark shifted frequency is well-defined over a wide range of pump powers and exhibits a smooth oscillating behavior decaying to $\widetilde\omega_a$.} This curve is in good agreement with the first-order predictions by a model resulting from time-averaging the Hamiltonian~\eqref{eq:shunted_displaced}
\begin{equation}\label{eq:av}\begin{split}
    \thH_{\text{av}} = &\, \hbar \widetilde{\omega}_a \, \tha^{\dag}\tha + %
        \hbar \widetilde{\omega}_b \, \thb^{\dag}\thb \\
    & - J_0(\xi) E_J \cos\left[\varphi_a^0\left(\tha + \tha^{\dag}\right) + \varphi_b^0 \left(\thb + \thb^{\dag}\right)\right],
\end{split}\end{equation}
where $J_0(\cdot)$ represents the Bessel function of the first kind. We note that the observed jump in the AC Stark shift of the un-shunted case appears at pump strengths much lower than the first oscillation of  this Bessel function. An experimental observation of such an oscillating behavior will prove a striking difference with the un-shunted case.

{This analysis indicates that we should also be able to  tune the strength of various types of nonlinear Hamiltonians such as the induced Kerr of the mode $\tha$~\cite{Kirchmair2013}. In Fig.~\ref{fig:extended}(b), we plot the Kerr strength calculated from Floquet simulations (blue dots). Such a simulation is performed by determining the first and the second excited Floquet states coupled to the ones in the limit cycle. Indeed, the Kerr strength is given by the difference of the transition frequencies for the first and the second excitation. Interestingly, we observe that the Kerr term vanishes for high enough powers. This ability in canceling the leading order nonlinear effects by merely tuning a pump power will be an extremely useful tool for circuit QED experiments~\cite{Frattini2017}. Furthermore, we plot the expected Kerr strength computed numerically from the time-averaged model~\eqref{eq:av} (orange curve). This is a first-order approximation of the Kerr effect and represents well its qualitative behavior.  In order to achieve a more precise approximation, we require to perform higher order rotating-wave approximations~\cite{QuantSys2015}.}

\section{Conclusion}\label{sec:conc}
In summary, we have investigated the non-linear dissipative dynamics of a Josephson circuit in the presence of strong off-resonant drives. Drive and dissipation are central ingredients of many recent parametric protocols to engineer various linear or nonlinear Hamiltonians in the context of circuit QED. Through the analysis of the steady state of the driven system coupled to a cold bath, we demonstrated that the transmon circuit, commonly used for such a purpose, \red{displays a structural instability in the exploitable range of pump powers. } Indeed, even for moderate pump powers and zero-temperature bath, this steady state is significantly mixed  and takes its support on  transmon states that are not confined in the cosine potential of the Josephson junction. {The transmon states are progressively transformed into states acting as those of  a free rotor, which do not induce any AC Stark shift of the oscillator. {In contrast,}  shunting the transmon circuit with an appropriate inductance \red{prevents the structural instability of the system} and considerably increases the purity of its states.}
Therefore the nonlinearity of the Josephson junction can be exploited over a wide range of pump strengths. In particular, the induced Kerr effect can be canceled out with high enough pump powers, \red{while maintaining other signatures of nonlinearity}. Finally, the Floquet type analysis performed in this paper can be extended to other similar problems, such as the study of  the dependence of the relaxation rate of a transmon qubit on the  dispersive readout strength~\cite{Sank2016,Mundhada-APS-2016}.

{\it Acknowledgements --}
We gratefully acknowledge useful discussions with Steven Girvin, Leonid Glazman, Manuel Houzet, Benjamin Huard, Pierre Rouchon and Alain Sarlette.
This research was supported by the ANR grant ENDURANCE, the EMERGENCES grant ENDURANCE of Ville de Paris, by ARO under Grant No. W911NF-14-1-0011, and by Inria’s DPEI under the TAQUILLA associated team.

\appendix
\section{Simulated models}\label{sec:models}
\subsection{Un-shunted transmon}\label{ssec:unshunted}
We start with the  Hamiltonian of the circuit shown in Fig.~\ref{fig:main}(a)
\begin{multline}
    \label{un-shunted-quantized-hamiltonian}
    \mathbf{H} = \hbar \omega_a\, \mathbf{a}^{\dagger} \mathbf{a} + 4 E_C \left(\mathbf{N} - N_g\right)^2 - E_J \cos(\bm{\theta}) \\
    + i \hbar g \left(\mathbf{N} - N_g\right)\left(\mathbf{a}^{\dagger} - \mathbf{a}\right) + i \hbar \mathcal{A}_p(t)\left(\mathbf{a}^{\dagger} - \mathbf{a}\right)
\end{multline}
where $\omega_a$ is the frequency of the bare harmonic oscillator (in absence of coupling to the transmon), $E_C$ and $E_J$ are the capacitive and Josephson energies of the transmon and $g$ is the coupling strength. The pump is described by $\mathcal{A}_p(t) = A_p \cos(\omega_p t)$ where $A_p$ is the pump amplitude and $\omega_p$ is the pump frequency. Here, $\mathbf{N}$ and $\cos(\bm{\theta})$ are the transmon mode operators corresponding to the number and transfer of Cooper pairs across the junction
\begin{equation}\left\{\begin{array}{l}
    \mathbf{N} = \sum_{N = -\infty}^{+\infty} \ket{N} \bra{N} \\
    \cos(\bm{\theta}) = \frac{1}{2} \sum_{N = -\infty}^{+\infty} \ket{N}\bra{N+1} + \text{h.c.} \\
\end{array}\right.\end{equation}
and $\mathbf{a}$ is the cavity annihilation operator. Also, $N_g$ is the offset charge of the superconducting island. We model the dissipation as a capacitive coupling of the cavity to the transmission line provided by~\eqref{eq:bath1}.

Let us displace the modes as $\widetilde{\mathbf{a}} = \mathbf{a} - \bar{a}(t)$ and $\widetilde{\bm{\theta}} = \bm{\theta} - \bar{\theta}(t)$ where
\begin{align*}
    \bar{a}(t) &= \frac{A_p}{2 i} \left[\frac{e^{i \omega_p t}}{\omega_a + \omega_p} + \frac{e^{-i\omega_p t}}{\omega_a - \omega_p}\right] \\
    \bar{\theta}(t) &= \frac{2 A_p g \omega_a}{\omega_p\left(\omega_a^2 - \omega_p^2\right)} \sin\left(\omega_p t\right) \quad \text{mod } (2\pi).
\end{align*}
Note that, here the displacement of $\bm{\theta}$ is equivalent to the application of a unitary given by $\mathbf{U}=\exp(i\bar\theta(t)\mathbf{N})$.

The Hamiltonian in the displaced frame is given by
\begin{multline}
    \widetilde{\mathbf{H}}(t) = \hbar\omega_a\, \widetilde{\mathbf{a}}^{\dagger} \widetilde{\mathbf{a}} + 4 E_C \left({\mathbf{N}} - N_g\right)^2 - E_J \cos(\bm{\widetilde{\theta}} + \xi \sin(\omega_p t )) \\+
     i \hbar g \left({\mathbf{N}} - N_g\right)\left(\widetilde{\mathbf{a}}^{\dagger} - \widetilde{\mathbf{a}}\right)
\end{multline}
where
$$
    \xi = \frac{2 A_p g \omega_a}{\omega_p\left(\omega_a^2 - \omega_p^2\right)}.
$$
This displacement brings the number of excitations in the harmonic oscillator close to zero. Additionally, it takes the pump drive into account as a drive on the superconducting phase of the transmon, inside the cosine term. These properties  make the numerical simulations tractable.

At this point, one should note that under this change of variables, the coupling to the bath \eqref{eq:bath1} is the same, using $\widetilde{\mathbf{a}}^{\dagger}$ and $\widetilde{\mathbf{a}}$ operators instead of $\mathbf{a}^{\dagger}$ and $\mathbf{a}$.

\subsection{Inductively shunted transmon}\label{shunted-transmon-quantization}

The  Hamiltonian of the circuit shown in Fig.~\ref{fig:main}(d) is given by
\begin{multline}
    \label{shunted-hamiltonian}
    \mathbf{H_{\text{shunt}}}(t) = \hbar\omega_a\mathbf{a}^\dag \mathbf{a} +
        4 E_C \mathbf{N}^2 + \frac{E_L}{2} \bm{\varphi}^2  - E_J \cos\bm{\varphi}\\
        + i\hbar g\, \mathbf{N}  (\mathbf{a}^\dag-\mathbf{a}) + i\hbar \mathcal{A}_p(t)(\mathbf{a}^\dag-\mathbf{a}).
\end{multline}
This is similar to the Hamiltonian of the previous subsection, except for the additional term corresponding to energy of the inductive shunt $E_L \bm{\varphi}^2/2$. Also, as a result of removing the superconducting island, and in contrast to the case of the previous subsection,  the phase $\bm{\varphi}$ is no more a compact variable and takes its values over entire $\mathbb{R}$. This is why we use a different notation from the un-shunted case: $\bm{\theta}$ stands for a phase defined in the compact interval $[0,2\pi]$ and $\bm{\varphi}$ is a phase defined over the entire $\mathbb{R}$.

We start by defining $\bb=(\bm{\varphi}+i\mathbf{N})/\sqrt{2}$. In the aim of diagonalizing the system and displacing it to take into account the drive, we perform, in order, a Bogoliubov transformation $\bU_{s1}$,   a beam-splitter type unitary $\bU_{\theta}$, a displacement of the frame $\bD$, and another Bogoliubov transformation $\bU_{s2}$ given by
\begin{align*}
\bU_{s1}&=\exp\left(\frac{\zeta}{2}(\bb^{\dag 2}-\bb^2)\right)\\
\bU_\theta&= \exp(\theta (\ba\bb^\dag-\ba^\dag\bb))\\
\bD&= \exp(\alpha^*(t)\ba-\alpha(t)\ba^\dag)\exp(\beta^*(t)\bb-\beta(t)\bb^\dag)\\
\bU_{s2}&=\exp\left(\frac{\zeta_a}{2}(\ba^{\dag 2}-\ba^2)\right)\exp\left(\frac{\zeta_b}{2}(\bb^{\dag 2}-\bb^2)\right).
\end{align*}
Here
\begin{align*}
\theta&=-\frac{1}{2}\arctan\left[\frac{2\hbar g \sqrt{2E_L\hbar\omega_a}}{(\hbar\omega_a)^2-8E_C E_L}\right],\\
\alpha(t)&=\frac{A_p\cos\theta}{\omega_p^2-\omega_a\omega_1}\left(\omega_p\sin(\omega_p t)+i\omega_a\cos(\omega_p t)\right),\\
\beta(t)&=\frac{A_p\sin\theta}{\omega_p^2-\omega_a\omega_2}\left(\omega_p\sin(\omega_p t)+i\omega_a\cos(\omega_p t)\right),\\
\zeta&=\log\left(\sqrt{\frac{E_L}{\hbar\omega_a}}\right),\\
\zeta_a&=\log\left(\sqrt[4]{\frac{\omega_a}{\omega_1}}\right),\quad
\zeta_b=\log\left(\sqrt[4]{\frac{\omega_a}{\omega_2}}\right),
\end{align*}
with
\begin{align*}
\omega_1&=\omega_a\cos^2\theta+\frac{8E_CE_L}{\hbar^2\omega_a}\sin^2\theta- g \sqrt{\frac{2 E_L}{\hbar \omega_a}} \sin(2\theta), \\
\omega_2 &= \omega_a \sin^2 \theta + \frac{8 E_C E_L}{\hbar^2 \omega_a} \cos^2\theta + g \sqrt{\frac{2 E_L}{\hbar \omega_a}}\sin(2\theta).
\end{align*}
This leads to a Hamiltonian given by
 \begin{multline}\label{eq:shunt-displaced}
     \mathbf{\widetilde{H}_{\text{shunt}}}(t)  =  \hbar \widetilde{\omega}_a \widetilde\ba^\dag\widetilde\ba+ \hbar \widetilde{\omega}_b \widetilde\bb^\dag\widetilde\bb\\
         - E_J \cos\left(\phi_a(\widetilde\ba^\dag+\widetilde\ba)+\phi_b(\widetilde\bb^\dag+\widetilde\bb)+\xi \sin(\omega_p t)\right)
\end{multline}
where
\begin{align*}
\widetilde{\omega}_a&=\sqrt{\omega_a\omega_1}, \quad\widetilde \omega_b=\sqrt{\omega_a\omega_2},\\
\phi_a&=-\sin(\theta)\sqrt{\frac{\hbar\omega_a}{2E_L}}\sqrt[4]{\frac{\omega_1}{\omega_a}},\quad\phi_b=\cos(\theta)\sqrt{\frac{\hbar\omega_a}{2E_L}}\sqrt[4]{\frac{\omega_2}{\omega_a}}, \\
 \xi&=A_p\omega_p\sin(2\theta)\sqrt{\frac{\hbar\omega_a}{2E_L}}\left(\frac{1}{\omega_p^2-\omega_a\omega_2}-\frac{1}{\omega_p^2-\omega_a\omega_1}\right).
\end{align*}
Under this change of variables, the coupling to the bath \eqref{eq:bath1} through the operator $i(\ba-\ba^\dag)$, is replaced by
$$
\mathbf{\nu}=i\cos(\theta)\sqrt[4]{\frac{\omega_1}{\omega_a}}(\widetilde \ba-\widetilde \ba^\dag)+i\sin(\theta)\sqrt[4]{\frac{\omega_2}{\omega_a}}(\widetilde \bb-\widetilde \bb^\dag).
$$
Therefore, the new system-bath coupling is given by
\begin{multline}\label{eq:bathshunt}
\mathbf{\widetilde H_{SB}}=\sum_{k}\hbar \omega_k\mathbf{c}^{\dagger}[\omega_k]\mathbf{c}[\omega_k]\\
- \hbar \Omega[\omega_k]\cos(\theta)\sqrt[4]{\frac{\omega_1}{\omega_a}}(\mathbf{\widetilde a}^{\dagger}  - \mathbf{\widetilde a})(\mathbf{c}^{\dagger}[\omega_k] - \mathbf{c}[\omega_k])\\
-\hbar \Omega[\omega_k]\sin(\theta)\sqrt[4]{\frac{\omega_2}{\omega_a}}(\mathbf{\widetilde b}^{\dagger}  - \mathbf{\widetilde b})(\mathbf{c}^{\dagger}[\omega_k] - \mathbf{c}[\omega_k]).
\end{multline}

\section{Floquet simulations}\label{sec:floquet}
\begin{figure}[h]
\begin{center}
\includegraphics[width=.4\textwidth]{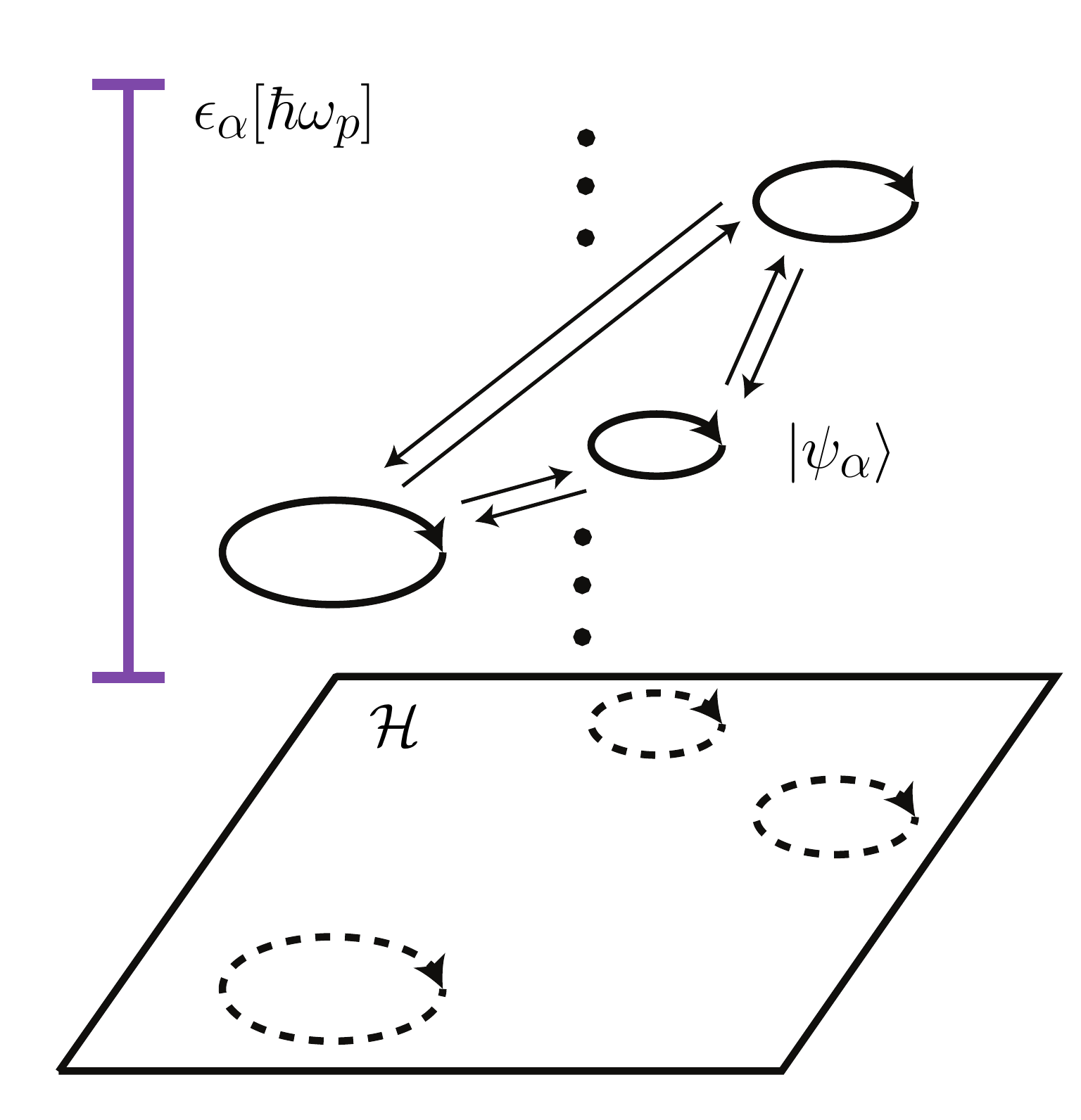}
\caption{Driven-dissipative quantum circuits and the Floquet-Markov theory. Floquet states $\left\{\ket{\Psi_\alpha(t)}\right\}_{\alpha}$ are periodic orbits of the driven system in its Hilbert space $\mathcal{H}$ . A quasi-energy $\epsilon_{\alpha}$ is associated to each Floquet state $\ket{\Psi_\alpha(t)}$. The set of quasi-energies is invariant under translation by multiples of $\hbar\omega_p$ (different Brillouin zones). Here we plot the Floquet states of the first Brillouin zone {(with quasi-energies defined modulo $\hbar\omega_p$ and denoted by $\epsilon_\alpha [\hbar\omega_p]$)} and their transitions due to the coupling to the bath. The steady state of the driven-dissipative system is given by a statistical mixture of these Floquet states, with populations inferred from an extension of the Fermi golden rule.}\label{fig:floquet}
\end{center}
\end{figure}
\subsection{Hamiltonian formulation of Floquet theory}
We consider here a system evolving under a time-periodic Hamiltonian $\widetilde H(t)$, of period $T = 2\pi / \omega_p$. Such a system can be efficiently simulated using the tools from the Floquet theory \cite[Section 2]{Grifoni1998}. In this section, we remind some of the basic elements of the Floquet theory that are required to understand the simulations of this paper. This material is borrowed and summarized from~\cite{Grifoni1998}.

The Schrödinger equation for this system is
\begin{equation}
    \label{floquet-schrodinger}
   i \hbar \frac{\partial }{\partial t}\ket{\widetilde{\Psi}(t)}
= \widetilde{\mathbf{H}}(t) \ket{\widetilde{\Psi}(t)}
    \end{equation}
where $\ket{\widetilde{\Psi}(t)}$ denotes the state of the system at time $t$. The Floquet theorem states that there exists solutions to \eqref{floquet-schrodinger} of the form
\begin{equation}
    \ket{\Psi_{\widetilde \alpha}(t)} = e^{-i \epsilon_{\widetilde\alpha} t / \hbar} \ket{\Phi_{\widetilde\alpha}(t)}
\end{equation}
where $\ket{\Phi_{\widetilde\alpha}}$ is called a \textit{Floquet mode} and is $T$-periodic in time and $\epsilon_{\widetilde\alpha}$ is a real-valued energy, called a \textit{quasi-energy}. In particular, we note that the set of quasi-energies is invariant under translation by multiples of $\hbar\omega_p$, as for any Floquet mode $\ket{\Phi_{\widetilde\alpha}(t)}$, the periodic wave-function $\exp(in\omega_p t)\ket{\Phi_{\widetilde\alpha}(t)}$ is also a Floquet mode. Therefore, the index $\widetilde\alpha$ corresponds to two indices $(\alpha,n)\in [-\hbar\omega_p/2,\hbar\omega_p/2[ \times\mathbb{Z}$ with $\epsilon_{\alpha,n}=\epsilon_{\alpha}+ n\omega_p$. Each value of $n$ here corresponds to a Brillouin zone. In these notes, we consider the first Brillouin zone $(\alpha,0)$ that we replace by $\alpha$ to simplify the notations.

A general approach to solve the above Schr\"odinger equation is to identify the Floquet modes and the associated quasi-energies. By decomposing the initial state as a superposition of the Floquet modes of the first Brillouin zone at time $t=0$, $\ket{\Psi(0)}=\sum_{\alpha}c_\alpha\ket{\Phi_\alpha(0)}$, the solution at time $t$ is given by
$$
\ket{\Psi(t)}=\sum_{\alpha}c_\alpha e^{-i\epsilon_\alpha t/\hbar}\ket{\Phi_\alpha(t)}.
$$
In order to identify the Floquet modes and the quasi-energies, we note that by applying the propagator $\widetilde{\mathbf{U}}(t + T, t)$ of~\eqref{floquet-schrodinger}, to a Floquet solution, we get
\begin{equation}
    \widetilde{\mathbf{U}}(t + T, t) \ket{\Phi_{\alpha}(t)} = e^{-i \epsilon_{\alpha} T / \hbar} \ket{\Phi_{\alpha}(T + t)}
\end{equation}
and in particular at $t=0$,
\begin{equation}
    \label{eq:floquet-numerical}
    \widetilde{\mathbf{U}}(T, 0) \ket{\Phi_{\alpha}(0)} = e^{-i \epsilon_{\alpha} T / \hbar} \ket{\Phi_{\alpha}(0)}.
\end{equation}

Equation \eqref{eq:floquet-numerical} can be used to numerically compute  the Floquet modes at $t = 0$ and their quasi-energies through the eigenstates and eigenvalues of $\widetilde{\mathbf{U}}(T, 0)$. Then, we can get the value of the Floquet mode at any later time using
\begin{equation}
    \ket{\Phi_{\alpha}(t)} = e^{i \epsilon_{\alpha} t / \hbar} \widetilde{\mathbf{U}}(t, 0) \ket{\Phi_{\alpha}(0)}.
\end{equation}

\subsection{Floquet-Markov approach for weak dissipation}
\label{floquet-markov}
\begin{figure}[h]
\includegraphics[width=.5\textwidth]{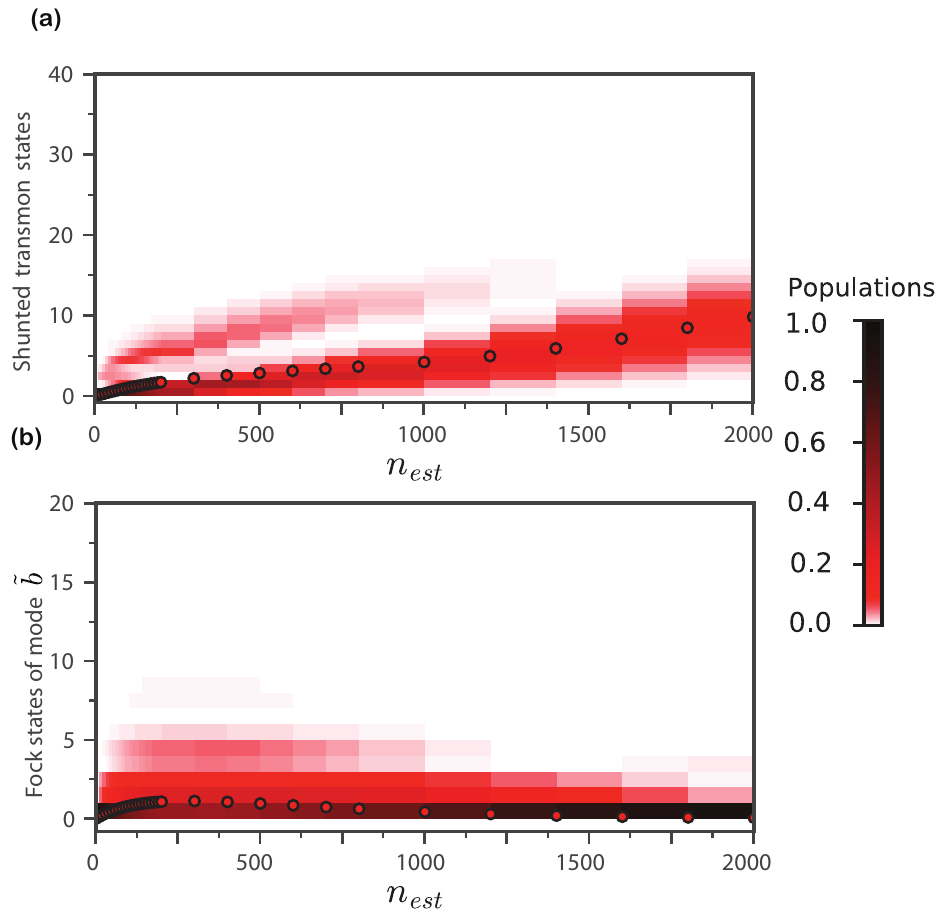}
\caption{Floquet-Markov simulations (asymptotic regime) of~\eqref{eq:shunt-displaced}-\eqref{eq:bathshunt} with $E_C/h=150$MHz, $E_J/h=6$GHz, $E_L/h=14$GHz, $g/2\pi=140$MHz, $\omega_a/2\pi=5.5$MHz, $\omega_p/2\pi=6$GHz. (a) The populations of the shunted transmon eigenstates $\ket{\nu_k}$ in the steady state $\rho_{ss}(0)$ as a function of pump power. (b) The populations of the $\widetilde \bb$-mode's Fock states in the same steady state.}\label{fig:displaced}
\end{figure}
\begin{figure*}[t]
\begin{center}
\includegraphics[width=\textwidth]{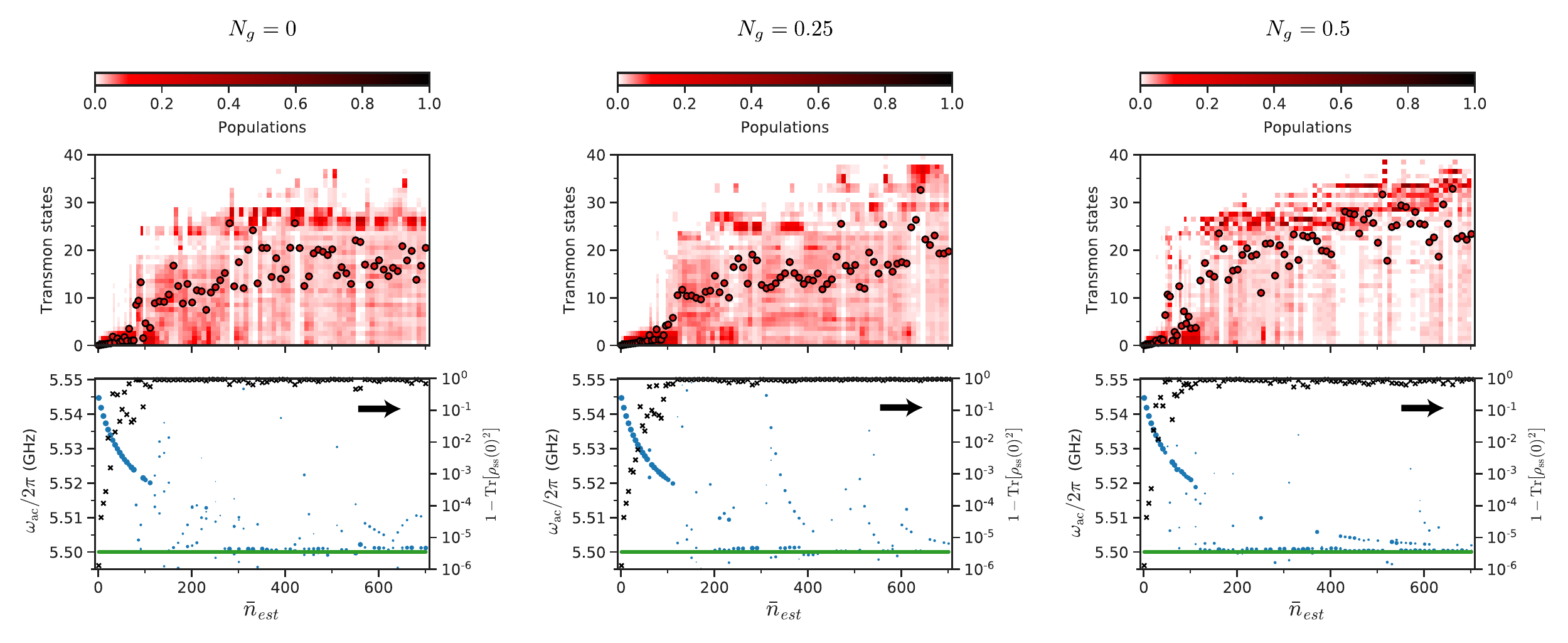}
\caption{Floquet-Markov simulations (in the asymptotic regime) of~\eqref{un-shunted-quantized-hamiltonian}-\eqref{eq:bath1}, using the same parameters as in Section~\ref{sec:unshunted} and three different values of $N_g$.  In the top figures, we plot  the populations of the transmon eigenstates in the steady state $\rho_{ss}(0)$ as a function of pump power. Also the red dots  indicate the average number of excitations in the transmon mode. In the bottom figures, we plot the AC Stark shifted frequencies of the oscillator  (blue dots) and the impurity of the steady state (black crosses, right axis) as a function of the pump power. As it can be seen, we observe no significant qualitative difference between the three cases. The steady state becomes rapidly very mixed and highly excited and the frequency is not well defined for $\bar n$ larger than 100.}\label{fig:ng}
\end{center}
\end{figure*}
The Floquet theory can be extended to take into account weak dissipations. Under the Floquet-Markov-Born approximation \cite[section 9]{Grifoni1998}, one can write a master equation in the basis of the Floquet modes of the first Brillouin zone:
\begin{equation}\label{floquet-steadystate}\begin{array}{l}
    \dot{\rho}_{\alpha \alpha}(t) = \sum_{\nu} \left[ L_{\alpha \nu} \rho_{\nu \nu}(t) - L_{\nu \alpha}\rho_{\alpha \alpha}(t) \right] \\
    \dot{\rho}_{\alpha \beta}(t) = - \frac{1}{2}\sum_{\nu} \left(L_{\nu\alpha} + L_{\nu \beta}\right) \rho_{\alpha \beta}(t), \quad \alpha \neq \beta \\
\end{array}
\end{equation}
where $\left(\rho_{\alpha \beta}\right)=\bra{\Phi_\alpha(t)}\rho\ket{\Phi_\beta(t)}$ are the components of the density matrix $\rho$. We have defined
\begin{equation}
    L_{\alpha \beta} = \sum_{k = -\infty}^{+\infty} \big(\gamma_{\alpha,\beta,k}+n_{\text{th}}(|\Delta_{\alpha,\beta,k}|) \left(\gamma_{\alpha,\beta,k} + \gamma_{\beta,\alpha,-k}\right)\big).
\end{equation}
Here,
\begin{equation}
    \gamma_{\alpha,\beta,k} = 2 \pi \Theta\left(\Delta_{\alpha \beta k}\right) J\left(\Delta_{\alpha,\beta,k}\right) \left|P_{\alpha \beta k}\right|^2
\end{equation}
where $\Theta$ is the Heaviside distribution, $\hbar \Delta_{\alpha,\beta,k} = \epsilon_{\beta} - \epsilon_{\alpha} + k \hbar \omega_p$ is a quasi-energy difference and $J(\omega)$ is the noise spectral function of the environmental coupling. The matrix elements, $P_{\alpha \beta k}$ are given by
\begin{equation}
    P_{\alpha \beta k} = \frac{i}{T} \int_0^T e^{-i k \omega_p t} \bra{\Phi_{\alpha}(t)} (\widetilde{\mathbf{a}} - \widetilde{\mathbf{a}}^{\dagger}) \ket{\Phi_{\beta}(t)} \,\mathrm{d}t.
\end{equation}
Finally, $n_{\text{th}}(\omega)=1/[\exp(\hbar\omega/k_B T)-1]$ is the thermal occupation of the bath at frequency $\omega$. In our simulations, we assume a zero temperature and therefore $n_{\text{th}}\equiv 0$.

Under some non-degeneracy assumptions (absence of resonance), the steady state of \eqref{floquet-steadystate} is diagonal in the Floquet modes basis. Moreover, the diagonal of this steady state density matrix can be numerically  computed, by solving the linear system
$    R p = 0$, where $\left(p_{\alpha}\right)_{\alpha} = \left(\rho_{\alpha \alpha}\right)_{\alpha}$ is the diagonal of the steady state density matrix and $\left(R_{\alpha \beta}\right)_{\alpha \beta} = \left(L_{\alpha \beta} - \delta_{\alpha\beta} \sum_{\nu} L_{\alpha \nu}\right)_{\alpha \beta}$ with $\delta_{\alpha \beta}$ the Kronecker delta.

\subsection{Numerical calculation of steady states}

The steady states in the simulations of Fig.~\ref{fig:main}b and e  have been numerically calculated following the above approach. We start by computing the Floquet modes and then reconstruct the stochastic transition matrix $R$. After calculating the steady state as a statistical mixture of the Floquet modes, we plot them in an appropriate basis of the Hilbert space. All our numerical simulations are run on a desktop workstation with an Intel Core i7-6700. We are running our simulations\cite{code} on a modified version of QuTiP 4.2.0~\cite{qutip,qutip2} under Python 3.5.2 and the plots have been produced using Matplotlib~\cite{Hunter:2007}. In the un-shunted case, and in the displaced frame provided in Appendix~\ref{ssec:unshunted}, we require a truncation of about 50 transmon states and 10 oscillator Fock states.  For the un-shunted case, we go to a frame provided in Appendix~\ref{shunted-transmon-quantization}. As a result of the stable behavior in this case, we require a smaller truncation of about 20 Fock states of the nonlinear mode and 10 Fock states of the linear one.

In Fig.~\ref{fig:main}(e), in order to put the system in a similar basis as in Fig.~\ref{fig:main}(b) for the un-shunted case, we plot the steady states in the shunted transmon basis after applying the inverse of the unitary transformations of Appendix~\ref{shunted-transmon-quantization}. Here, in Fig~\ref{fig:displaced}, we provide this steady state over an extended span of pump powers, and both in the shunted transmon basis $\ket{\nu_k}$ of $4E_C\bn^2+E_L\bph^2/2-E_J \cos(\bph)$, and in the distorted, rotated and displaced frame $\widetilde \ba$ and $\widetilde\bb$, of Appendix~\ref{shunted-transmon-quantization}. We see that in this second frame, the steady state remains very close to the ground state of the mode $\widetilde\bb$, for all values of the pump power. This statement is also true for the mode $\widetilde\ba$.

\begin{figure}[t]
\begin{center}
\includegraphics[width=.49\textwidth]{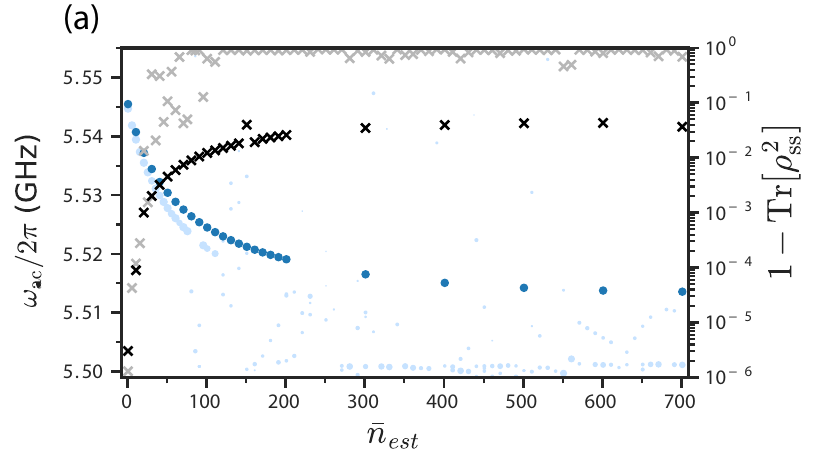}\\
\hspace{-.5cm}\includegraphics[width=.41\textwidth]{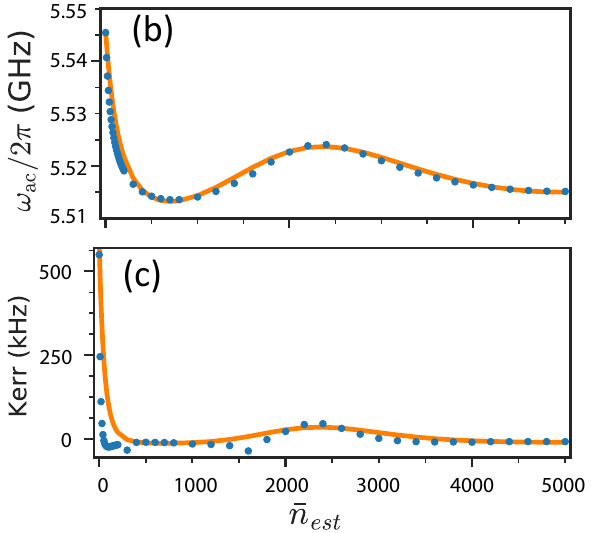}
\caption{Floquet-Markov simulations of the shunted transmon with  parameters  $E_C/h=450$ MHz, $E_J/h= 2.22$ GHz, $E_L/h=4.44$ GHz, $g/2\pi=245$ MHz, $\omega_a/2\pi=5.5$ GHz, $\omega_p/2\pi=6$ GHz. (a) Blue dots correspond to the AC Stark shifted frequencies of the oscillator as a function of the pump power. For comparison, we have also reproduced the results for the un-shunted case with the parameters of Fig.~\ref{fig:main}. These are plotted as pale blue dots (left axis). We also have plotted the impurity of the steady state (black crosses, right axis) versus the corresponding results for  the un-shunted case (gray crosses, right axis). (b) We plot the AC Stark shifted frequency with the shunted transmon over an extended range of pump powers. The blue dots correspond to the Floquet simulation results and the orange curve indicates the expected values from a time-averaged model. This is to be compared to Fig.~\ref{fig:extended}(a).(c) Induced Kerr strength over the same extended range (to be compared with Fig.~\ref{fig:extended}(b)).  }\label{fig:nonlin}
\end{center}
\end{figure}

\subsection{Computing AC-Stark shifts}
\label{stark-shift}
We are interested in the resonance frequency of the driven system with Hamiltonian $\widetilde{H}(t)$, close to the oscillator's bare frequency. Experimentally, we can find such a resonance frequency by sweeping the frequency of a very weak probe drive around the  oscillator's frequency~\cite{Lescanne2018}. We model this weak probe as a small perturbative Hamiltonian $i\hbar\varepsilon(t) \left(\widetilde{\mathbf{a}}^{\dagger} - \widetilde{\mathbf{a}}\right)$.

As shown in previous subsections, the system converges asymptotically to a limit cycle given by a statistical mixture of Floquet states:
$$
\rho_{ss}(t)=\sum_{\alpha} p_\alpha \ket{\Phi_{\alpha}(t)}\bra{\Phi_\alpha(t)}.
$$
Initializing the system at one of the Floquet modes $\ket{\Phi_\alpha}$ populated in the steady state, let us focus on the solution of the Schr\"odinger equation in the presence of the weak probe. We consider this solution at the lowest order in the amplitude of the probe field. The Schrödinger equation in this case is
\begin{equation}\label{eq:Sch}
    \frac{\partial}{\partial t} \ket{\Psi(t)}= - \frac{i}{\hbar} \mathbf{H}_\varepsilon(t) \ket{\Psi(t)},~ \ket{\Psi(0)}=\ket{\Phi_\alpha(0)}
\end{equation}
where $\mathbf{H}_\varepsilon(t) = \widetilde{\mathbf{H}}(t) + i \varepsilon(t) \left(\widetilde{\mathbf{a}}^{\dagger} - \widetilde{\mathbf{a}}\right)$.

First, let us introduce the propagation operator $\widetilde{\mathbf{U}}(t, 0)$ associated with the $\widetilde{\mathbf{H}}(t)$ Hamiltonian,
\begin{equation*}
    \frac{\partial \widetilde{\mathbf{U}}(t, 0)}{\partial t} = - \frac{i}{\hbar} \widetilde{\mathbf{H}}(t) \widetilde{\mathbf{U}}(t, 0),\qquad \widetilde{\mathbf{U}}(0, 0)=\mathbf{I}.
\end{equation*}
The solution of~\eqref{eq:Sch} is given by
\begin{align}
    \label{perturbative-evolution}
    &\ket{\Psi(t)} = \widetilde{\mathbf{U}}(t, 0)\ket{\Phi_{\alpha}(0)}\notag \\
    &~~+ \frac{1}{\hbar}\, \widetilde{\mathbf{U}}(t, 0) \int_0^t \varepsilon(s) \widetilde{\mathbf{U}}(s, 0)^{\dagger} \left(\widetilde{\mathbf{a}}^{\dagger} - \widetilde{\mathbf{a}}\right) \widetilde{\mathbf{U}}(s,0)\ket{\Phi_\alpha(0)}\,\mathrm{d}s \notag\\
    & =e^{-i\epsilon_\alpha t/\hbar}\ket{\Phi_\alpha(t)}\\
    &~~+ \frac{1}{\hbar}\, \widetilde{\mathbf{U}}(t, 0) \int_0^t \varepsilon(s)e^{-i\epsilon_\alpha s/\hbar} \widetilde{\mathbf{U}}(s, 0)^{\dagger} \left(\widetilde{\mathbf{a}}^{\dagger} - \widetilde{\mathbf{a}}\right) \ket{\Phi_\alpha(s)}\,\mathrm{d}s.\notag
    \end{align}
Let us now focus on the overlap of $\ket{\Psi(t)}$ with  other Floquet modes $\ket{\Phi_{\beta}(t)}$. We have from\eqref{perturbative-evolution},\small
\begin{multline}
    \braket{\Phi_{\beta}(t)}{\Psi(t)} = e^{-i\epsilon_\alpha t/\hbar} \braket{\Phi_{\beta}(t)}{\Phi_{\alpha}(t)} \\+ \frac{1}{\hbar} \bra{\Phi_{\beta}(t)} \widetilde{\mathbf{U}}(t, 0) \int_0^t \varepsilon(s)e^{-i\epsilon_\alpha s/\hbar} \widetilde{\mathbf{U}}(s, 0)^{\dagger} \left(\widetilde{\mathbf{a}}^{\dagger} - \widetilde{\mathbf{a}}\right) \ket{\Phi_{\alpha}(s)}\,\mathrm{d}s
\end{multline}\normalsize
that is\small
\begin{align*}
    &\braket{\Phi_{\beta}(t)}{\Psi(t)} = e^{-i\epsilon_\alpha t/\hbar}\braket{\Phi_{\beta}(t)}{\Phi_{\alpha}(t)} \\&~~ + \frac{1}{\hbar} e^{-i\epsilon_\beta t/\hbar} \int_0^t \varepsilon(s)e^{i(\epsilon_\beta-\epsilon_\alpha)s/\hbar}   \bra{\Phi_{\beta}(s)} \left(\widetilde{\mathbf{a}}^{\dagger} - \widetilde{\mathbf{a}} \right) \ket{\Phi_{\alpha}(s)}\,\mathrm{d}s\\
    &~~ =e^{-i\epsilon_\alpha t/\hbar}\braket{\Phi_{\beta}(t)}{\Phi_{\alpha}(t)} \\&~~ -\frac{i}{\hbar} e^{-i\epsilon_\beta t/\hbar}  \sum_k \int_0^t \varepsilon(s) e^{i \Delta_{\alpha,\beta ,k}s} P_{\beta,\alpha,k} ds
\end{align*}\normalsize
To induce a transition in the system between the  Floquet modes $\ket{\Phi_\alpha}$ and $\ket{\Phi_\beta}$, one needs  the frequency of the probe drive $\varepsilon(t)$ to match one of the frequencies $\Delta_{\alpha, \beta, k}$, and furthermore that the associated matrix element $P_{\beta,\alpha,k}$ is non-zero. Moreover, the transition rate is proportional to both the population of the initial Floquet mode  $\ket{\Phi_\alpha}$ in the steady state $\rho_{ss}$ given by $p_\alpha$, and the matrix element $P_{\beta,\alpha,k}$. In Figs.~\ref{fig:main}c and f, we have plotted the predominant transition frequencies at each pump power.

\subsection{un-shunted transmon and charge offset}
In this subsection, we focus on the un-shunted case and we study the effect of the  charge offset $N_g$. In the simulations of Section~\ref{sec:unshunted}, we have taken $N_g=0$. As we see in Fig.~\ref{fig:ng}, the choice of $N_g$ in the Hamiltonian~\eqref{un-shunted-quantized-hamiltonian} does not have any significant effect on the qualitative behavior of the system in the steady state.

\subsection{Choice of parameters for shunted transmon}

\begin{figure}[t]
\begin{center}
\includegraphics[width=.5\textwidth]{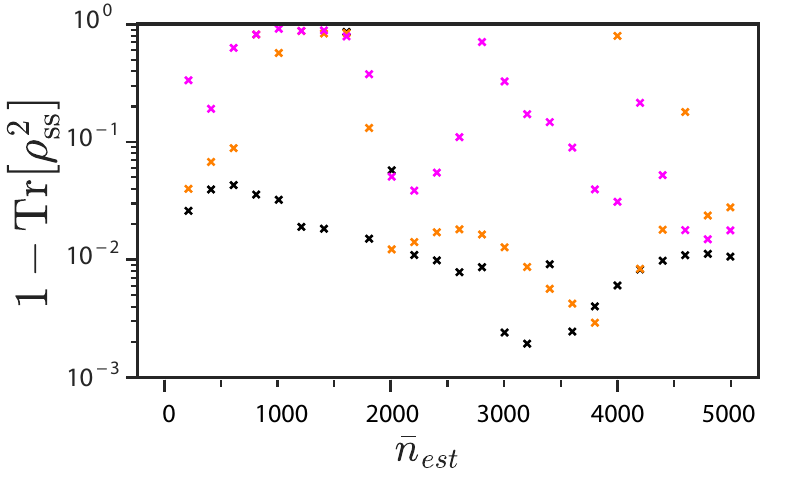}
\caption{Impurity of the steady state as a function of pump power. We use the same parameters as in Fig.~\ref{fig:nonlin}, except for $E_J$ and $E_L$. While the sum $(E_J+E_L)/h=6.66$ GHz is fixed, we take 3 different choices for their ratio $r=E_L/E_J$. The black crosses correspond to $r=2$ ($E_L/h=4.44$ GHz and $E_J/h=2.22$ GHz), the orange ones to $r=1.5$ ($E_L/h=4$ GHz and $E_J/h=2.22$ GHz), and the magenta ones to $r=1$ ($E_L/h=E_J/h=3.33$ GHz).}\label{fig:ELEJ}
\end{center}
\end{figure}

The simulations of Section~\ref{sec:shunted} have been performed with the same parameters as in the un-shunted case, except for the Josephson energy that has been taken to be $E_J/h=6$ GHz and the addition of $E_L/h= 14$ GHz. Noting that the sum of these two energies correspond to the Josephson energy in the un-shunted case, this choice allows to keep the bare frequency of the transmon mode the same. This, however, comes at the expense of diluting the nonlinearity of the transmon mode. Indeed, the anharmonicity of the shunted transmon mode is given by 37 MHz, to be compared to 143 MHz in the un-shunted case.  In the same way the induced Kerr on the cavity of 306 kHz is  weaker than 655 kHz, for the un-shunted case. The shallower slope of the AC Stark shift in Fig.~\ref{fig:main}(f) (with respect to Fig.~\ref{fig:main}(c)) can be explained through this difference.

Using a different set of parameters, one can achieve similar nonlinearities for the shunted transmon. For instance, by choosing $E_C/h=450$ MHz, $E_J/h= 2.22$ GHz, $E_L/h=4.44$ GHz, $g/2\pi=245$ MHz, $\omega_a=5.5$ GHz, we achieve similar frequencies and nonlinearities to the shunted case. More precisely, in the absence of the pump, we find the cavity frequency to be $5.545$ GHz, the qubit frequency $4.7$ GHz, the qubit anharmonicity 123 MHz, the induced cavity Kerr of 600 kHz, and a cross Kerr between the qubit and the cavity of $15.5$ MHz. These parameters for the un-shunted case are respectively given by $5.545$ GHz,  $4.691$ GHz, $143$ MHz, $655$ kHz, $17.3$ MHz. In Fig.~\ref{fig:nonlin}(a), we plot and compare the shifted cavity frequencies in the shunted and un-shunted case (blue dots, left axis). The slope near $\bar n_{\text{est}}=0$ of the variation of frequency vs photon number $\bar n_{\text{est}}$ is now  very close  to that of the un-shunted case. We also plot the impurity of the steady state in both cases versus the pump power (black and gray crosses, right axis). One clearly observes a much purer and smoother behavior for the shunted case with respect to the un-shunted one. In Fig.~\ref{fig:nonlin}(b) and (c), we plot the shifted cavity frequency and induced Kerr effect over a larger range of pump powers for the shunted case with these new parameters. We observe a behavior similar to that shown in panels (a) and (b) of Fig.~\ref{fig:extended}. As a result of the increased non-linearity, the range of the values taken by the Kerr strength is twice larger than in the simulations of the shunted case with the parameters in Section~\ref{sec:shunted}.

In these simulations, similar to the previous set of parameters, we have chosen a ratio between $E_L$ and $E_J$ of about 2. Noting that a large ratio between $E_L$ and $E_J$ leads to the dilution of the Josephson junctions non-linearity, one may consider the possibility of choosing a smaller ratio. We will see however that, this comes at the expense of losing the purity of the steady state and therefore getting closer to the asymptotic behavior in the un-shunted case. In order to illustrate this, we perform numerical simulations with the same parameters as in Fig.~\ref{fig:nonlin}, except for $E_J$ and $E_L$. Indeed, we fix their sum $(E_J+E_L)/h=6.66$ GHz, and we let vary the ratio between them. In Fig.~\ref{fig:ELEJ}, we provide the impurity of the steady state as a function of the pump power for 3 different choices of the ratio $r=E_L/E_J$. As it can be seen a ratio of 2, as chosen in this paper, ensures globally a purer steady state  and this purity is lost for smaller ratios.

%
%
%

\bibliography{main.bib}

\end{document}